\documentclass[apj]{emulateapj}

\usepackage{url}
\usepackage{wasysym}

\begin{document}


\title{A Comparison Between Physics-based and Polytropic MHD Models for Stellar Coronae and Stellar Winds of Solar Analogs}


\author{O. Cohen\altaffilmark{1,2}}

\altaffiltext{1}{Lowell Center for Space Science and Technology, University of Massachusetts, Lowell, MA 01854, USA}
\altaffiltext{2}{Harvard-Smithsonian Center for Astrophysics, 60 Garden St. Cambridge, MA 02138, USA}

\begin{abstract}

The development of the Zeeman-Doppler Imagine (ZDI) technique has provided synoptic observations of surface magnetic fields of low-mass stars. This led the Stellar Astrophysics community to adopt modeling techniques that have been used in solar physics using solar magnetograms. However, many of these techniques have been neglected by the solar community due to their failure to reproduce solar observations. Nevertheless, some of these techniques are still used to simulate the coronae and winds of solar analogs. Here we present a comparative study between two MHD models for the solar corona and solar wind. The first type of model is a polytropic wind  model, and the second is the physics-based AWSOM model. We show that while the AWSOM model consistently reproduces many solar observations, the polytropic model fails to reproduce many of these observations and in the cases it does, its solutions are  unphysical. Our recommendation is that polytropic models, which are used to estimate mass-loss rates and other parameters of solar analogs, must be first calibrated with solar observations. Alternatively, these models can be calibrated with models that capture more detailed physics of the solar corona (such as the AWSOM model), and that can reproduce solar observations in a consistent manner. Without such a calibration, the results of the polytropic models cannot be validated, but they can be wrongly used by others. 

\end{abstract}



\section{INTRODUCTION}
\label{sec:Intro}

The Zeeman-Doppler Imagine (ZDI) technique \citep{DonatiSemel90} has provided, for the first time, synoptic observations of surface magnetic fields of low-mass stars (mostly M, K, and G). Despite of some criticism on the uncertainty of these stellar "magnetograms" \citep[e.g.,][]{Kochukhov10,Reiners12}, the growth in availability of stellar magnetograms \citep[see][for a summary of these observations]{Vidotto14b} has led the stellar astrophysics community to adopt techniques, which were used by the solar community to extrapolate the three-dimensional coronal magnetic field. In particular, ZDI data has been used to drive three-dimensional models for stellar coronae and stellar winds (in a similar manner that solar magnetograms are used to drive models for the solar corona and solar wind). 

The first approach, which was adopted from solar physics is the so-called "Potential Field Source Surface" (PFSS) method \citep{altschulernewkirk69}. In the PFSS model, the three-dimensional magnetic field is assumed to be static (i.e., there is no forcing on the field by electric currents) and as such, the field can be described as a gradient of a scalar potential. This scalar potential can be obtained by solving the Laplace's equation, assuming that the field is purely radial above a spherical surface - the "Source Surface", which is set at a random distance. The potentiality of the solar coronal field has been debated for some time among the solar community \citep[see e.g.,][]{Riley06}. In particular, the validity of a spherical source surface and the choice of its distance have been challenged frequently \citep[see][for recent review on the PFSS issues]{Cohen15}. Currently, the PFSS method is mainly used to calculate the initial state of the magnetic field in Magnetohydrodynamic (MHD) models or to obtain a tentative description of the coronal field. Due to its limitations, the solar community is currently transitioning from using the PFSS method to more sophisticated field extrapolation methods, such as the non-linear force free technique \citep[see review by][]{Wiegelmann12}. In contrast, the PFSS method is used more frequently to extrapolate stellar coronal fields based on the ZDI maps \citep[see e.g.,][]{Jardine99}

Following the pioneering coronal modeling work by \cite{PneumanKopp71}, modern, multi-dimensional MHD models for the solar corona have been first applied a thermally driven, polytropic Parker wind \citep{Parker58} on top a a potential field \citep{Linker90,Usmanov93,LinkerMikic95}. The Parker wind solution is obtained analytically assuming a (nearly) isothermal radially expanding flow. A steady-state is obtained when a {\it pressure balance} between the Parker wind and the potential field is achieved. Despite of spacecraft conformation of the existence of super-Alfv\'enic solar wind, the Parker wind and its implementation in these older MHD models have failed to reproduce the observation. In particular, they could not produce the observed fast solar wind. Thus, we should account for additional acceleration missing from the Parker model.

\cite{Holzer77} and \cite{Usmanov00} have pointed out that in order to properly reproduce the solar corona and the solar wind, MHD models need to include additional momentum and energy terms beyond the set of ideal MHD equation. While the physical interpretation of these terms is still under debate, these terms account for the observed coronal heating and wind acceleration. In an intermediate stage, the polytropic models have been extended to use varying polytropic index, $\gamma$, as a function of some local properties of the gas \citep[e.g.,][]{Roussev03,Cohen07,Feng10,Jacobs11}. In particular, models were developed to relate the local value of $\gamma$ to the observed empirical relation between the solar wind speed and the magnetic flux tube expansion geometry \citep{WangSheeley90,McComas07}. These models succesfully reproduced the observed solar wind. However, due to the lack of detailed thermodynamics, they did not provide a good agreement with observations of the solar corona itself (i.e., comparison with EUV/X-ray line-of-sight observations). Recent MHD models for the solar corona and solar wind adopt a physics-based approach and include additional energy and momentum sources, as well as thermodynamic terms (heating and cooling), and multi-fluid approach to include the separation between ions and electrons \citep[e.g., most recent models by][]{Downs14,Lionello14,Vanderholst14,Usmanov16}. 

Despite of this evolution in modeling the solar corona and wind (and perhaps due to some possible disconnection by the Astrophysics community from solar physics literature) recent models for stellar coronae of solar analogs have returned to use the spherical polytropic Parker wind imposed on a potential magnetic field. These models have used available ZDI maps or low-order idealized magnetic fields to study and scale stellar mass-loss rates and stellar spindown with stellar age and rotation periods, as well as to characterize the coronae and winds of planet-hosting stellar systems \citep[e.g.,][]{Matt12,Vidotto14b,Matsakos15,Reville15,Strugarek15}, while they were not calibrated by solar input parameters and solar magnetogram data. Instead, some of the models have adopted an idealized "fiducial" solar case, which is represented by solar mass, radius, rotation period, and a dipole magnetic field.

The goal of this paper is to demonstrate that the polytropic Parker wind imposed on a potential field struggles to reproduce solar observations. We use high-resolution solar magnetograms to drive both a polytropic Parker wind MHD model, and a physics-based MHD model for the solar corona and solar wind, and compare the output from each model with solar observations. In the next section we describe the models and the chosen input parameters, we present and discuss the results in Section~\ref{sec:Results}, and conclude our findings in Section~\ref{sec:Conclusions}.  

\section{Model Description}
\label{sec:Models}

In this study, we compare the solutions of two types of MHD models, both performed using the {\it BATS-R-US} MHD code \citep{powell99, Toth12}. 

\subsection{Thermally-driven Polytropic MHD Model}
\label{sec:ParkerModel}

In the first setting of {\it BATS-R-US}, the set of ideal MHD equations is solved {\it with no source terms in the momentum and energy equations} assuming a polytropic gas, where the value of $\gamma$ is close to unity (nearly isothermal gas). This solution provides some acceleration of the wind due to the pressure gradient between the solar surface and space, where the chosen coronal temperature determines the amount of gas expansion. This model does not account for any thermodynamic processes in the stellar corona except for the prescribe temperature of the gas and the adiabatic expansion. In other words, the corona in this model is assumed to be already heated, and the equations are relaxed to steady-state (when a pressure balance is achieved). In this polytropic setting, we use a prescribed, non-uniform Cartesian grid with a varying grid size ranging from $\Delta x = 0.02R_\odot$ near the inner boundary to $\Delta x=0.5R_\odot$ at the outer parts of the domain, which extends to $24R_\odot$.

\subsection{The AWSOM Model}
\label{sec:AWSOMModel}

In the second setting of {\it BATS-R-US}, we use the recently developed Alfv\'en Wave Solar Model (AWSOM) \citep{Sokolov13,Vanderholst14}. This model assumes a physics-based, self-consistent coronal heating and wind acceleration by Alfv\'en waves. Thus, it introduces additional energy and momentum terms that incorporate these physical processes. The energy spectrum of the Alfv\'en waves is calculated by assuming a turbulent cascade between two, counter-propagating waves along the magnetic field lines, where two additional equations are introduced for these two Alfv\'en waves. From these two equations, the total energy dissipation and the Alfv\'en waves pressure gradient are then added to the MHD momentum and energy equations. 

The AWSOM model accounts for thermodynamic and radiative transfer processes, such as electron heat conduction and radiative cooling, and could also be run in two-temperature mode, where the electrons and ions are decoupled (not used here). Unlike the polytropic model, at which the inner boundary is set at the coronal base, the inner boundary of AWSOM is set at the chromosphere. Finally, the Poynting Flux, which is specified at the base of model, is formulated so that it depends only on the stellar radius square, assuming the observed relation between the unsigned magnetic flux and the X-ray flux \citep{Pevtsov03}. This feature provides a built-in scaling from the model's original parametrization for the Sun to other, Sun-like stars. 

In the AWSOM setting, we use a spherical grid, which is stretched in the $r$ coordinate, with the smallest grid size being $\Delta r=0.015R_\odot$ near the inner boundary, and the largest grid size being $\Delta r=0.65$  near the outer boundary. The angular resolution is about two degrees.

\subsection{Models Parameters}
\label{sec:ModelParameters}

\begin{figure*}[h!]
\centering
\includegraphics[width=2.5in]{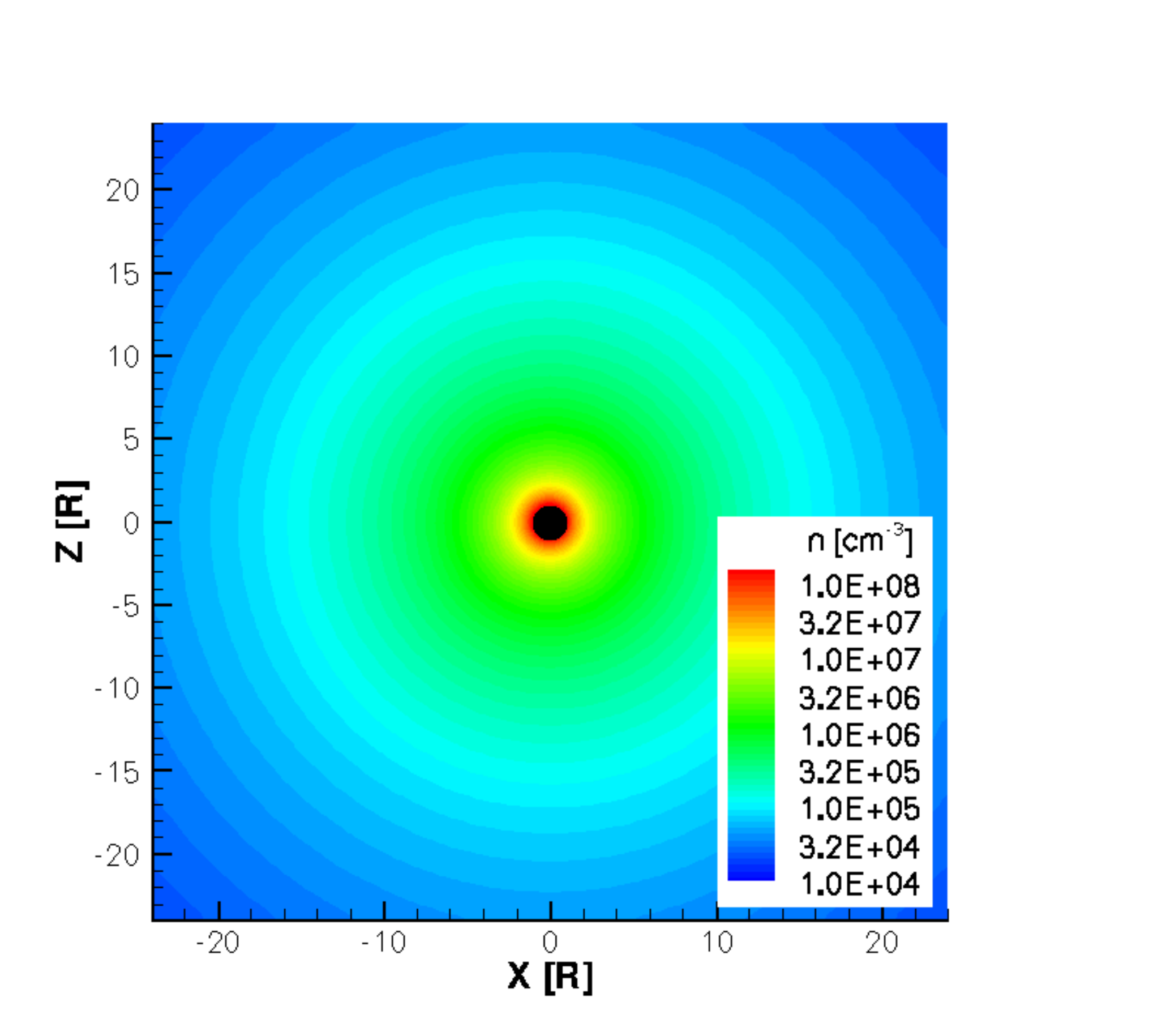}
\includegraphics[width=2.5in]{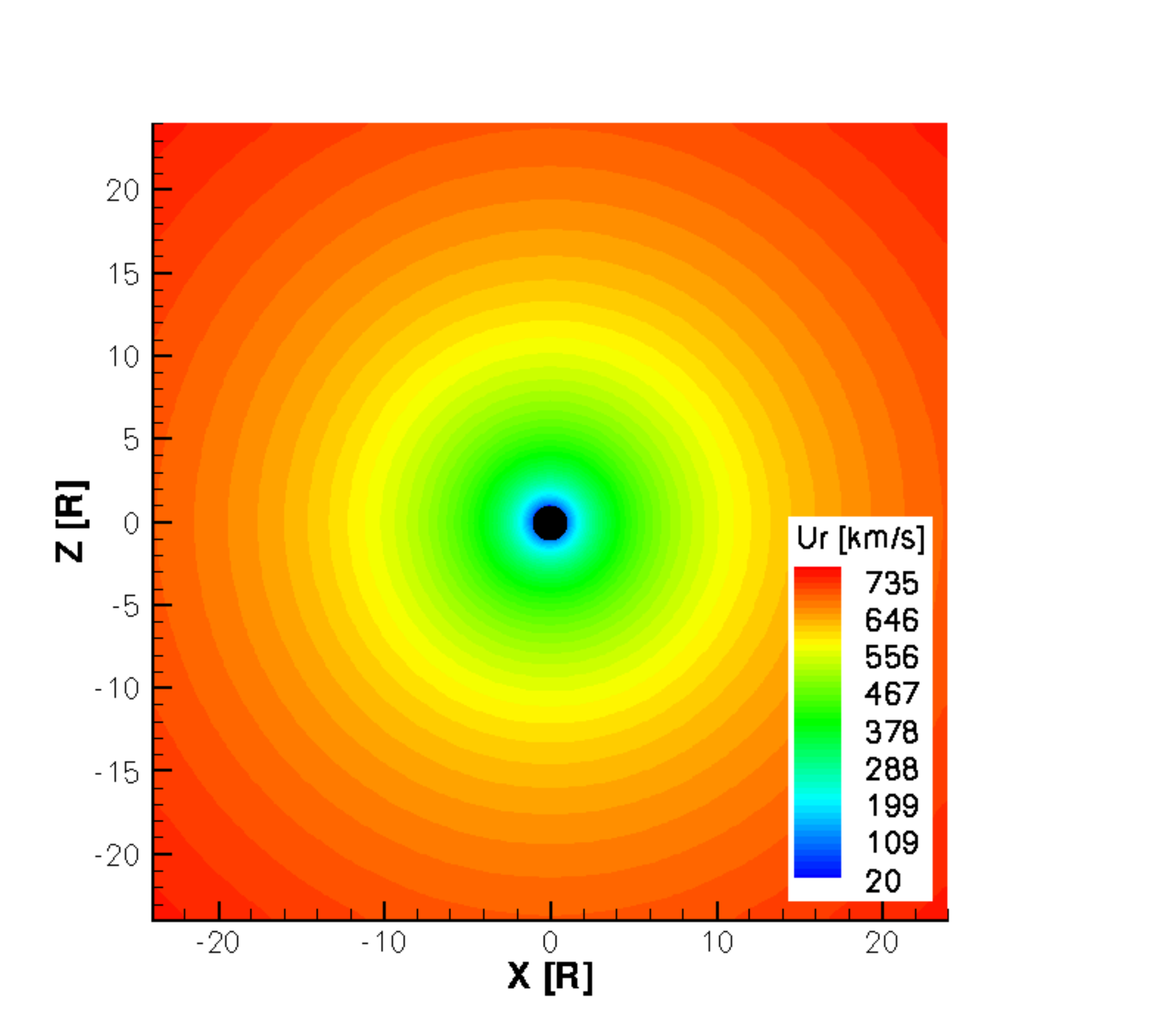}
\caption{Initial distribution of the coronal number density (left), and the Parker wind (with T=3MK) used for all simulated cases. The initial plasma temperature is $T_p=1.5$MK everywhere.}
\label{fig:f1}
\end{figure*}

In order to keep the two models as consistent which each other as possible we set both models with the same, magnetogram input data, the same initial conditions for the three-dimensional magnetic field (a PFSS extrapolation), the same initial density structure, and the same initial condition for the wind speed (a Parker solution with $T=3MK$). We also specify the boundary conditions to be the same in both models, with the coronal number density, $n=10^8\;[cm^{-3}]$, and the coronal temperature, $T=3MK$. We use this high coronal temperature to obtain un upper limit of the wind speed in the polytropic model. Figure~\ref{fig:f1} shows the initial conditions for the number density and wind speed. 

Despite of the different grid geometries used in the two settings, we trust that the comparison presented here is valid due to the fact that we are interested in comparing the overall wind acceleration via global properties, such as the maximum wind velocity and total mass-loss rate, and the fact that the grid size near both the inner and outer boundaries is comparable in both settings. For the Polytropic setting, we test three cases with $\gamma=1.01,1.05,$ and $1.1$. We also test how the coronal base density affects the solution by performing one case with $\gamma=1.05$ and $n=10^7\;[cm^{-3}]$.

\subsection{Input Data}
\label{sec:InputData}

In order to drive the two models, we use high-resolution Michelson Doppler Imager (MDI) solar magnetograms obtained from the Stanford Magnetogram repository (\url{http://hmi.stanford.edu/data/synoptic.html}). The magnetogram input data is used in the form of spherical harmonics coefficient list calculated up to $n=90$ (this resolution enables to resolve active regions on the Sun). We perform simulations for solar minimum period (quiet Sun) using a  magnetogram for Carrington Rotation (CR) 1916 (November-December 1996), and for solar maximum period (active Sun) using a magnetogram for CR 1962 (April-May 2000). In addition, we run the different model cases using a ZDI map of HD189733, which is reproduced from the data published in \cite{Fares10}. While this is a low-resolution magnetogram, we still extrapolate it up to $n=90$ for constancy, and in order to maintain high-resolution grid for the PFSS extrapolation since this resolution is determined by the order of harmonics.


\section{RESULTS}
\label{sec:Results}

\begin{figure*}[h!]
\centering
\includegraphics[width=5.5in]{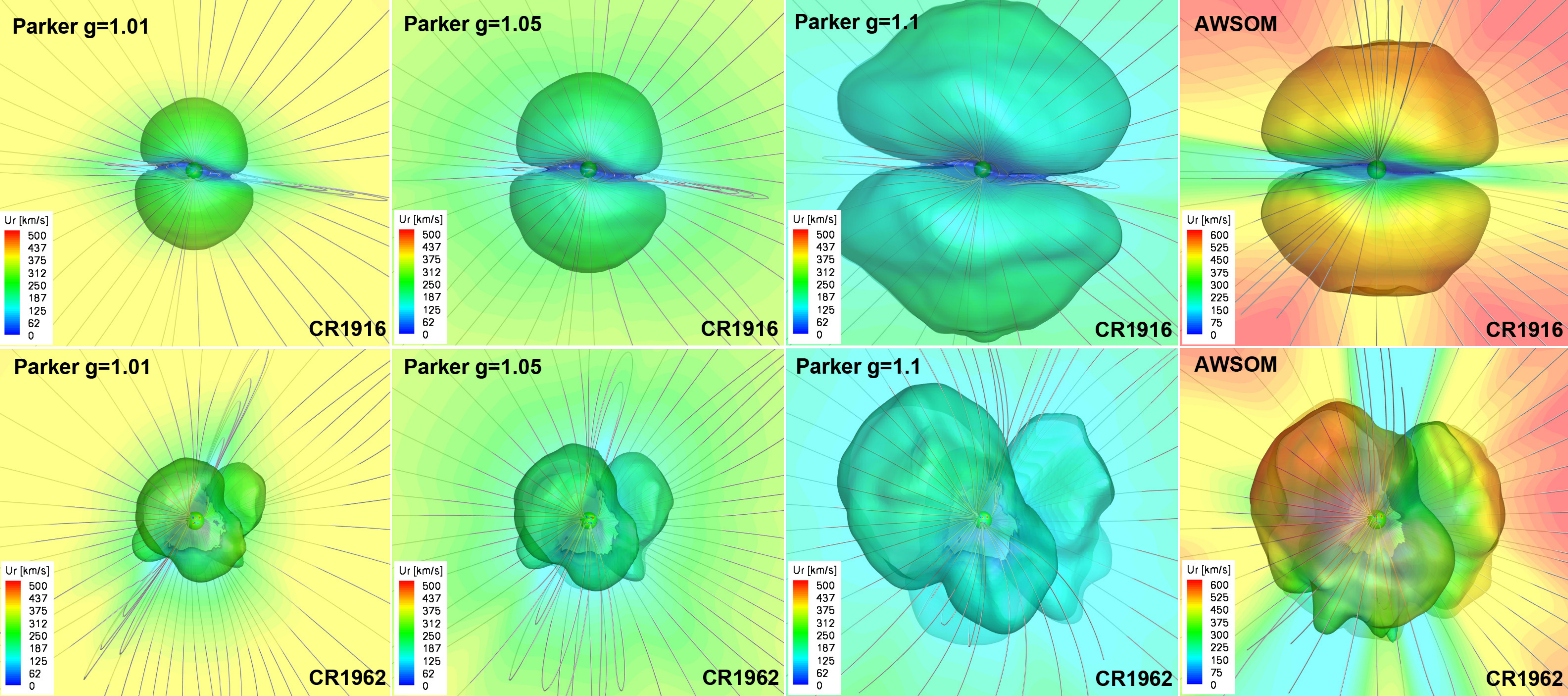}\\
\includegraphics[width=5.5in]{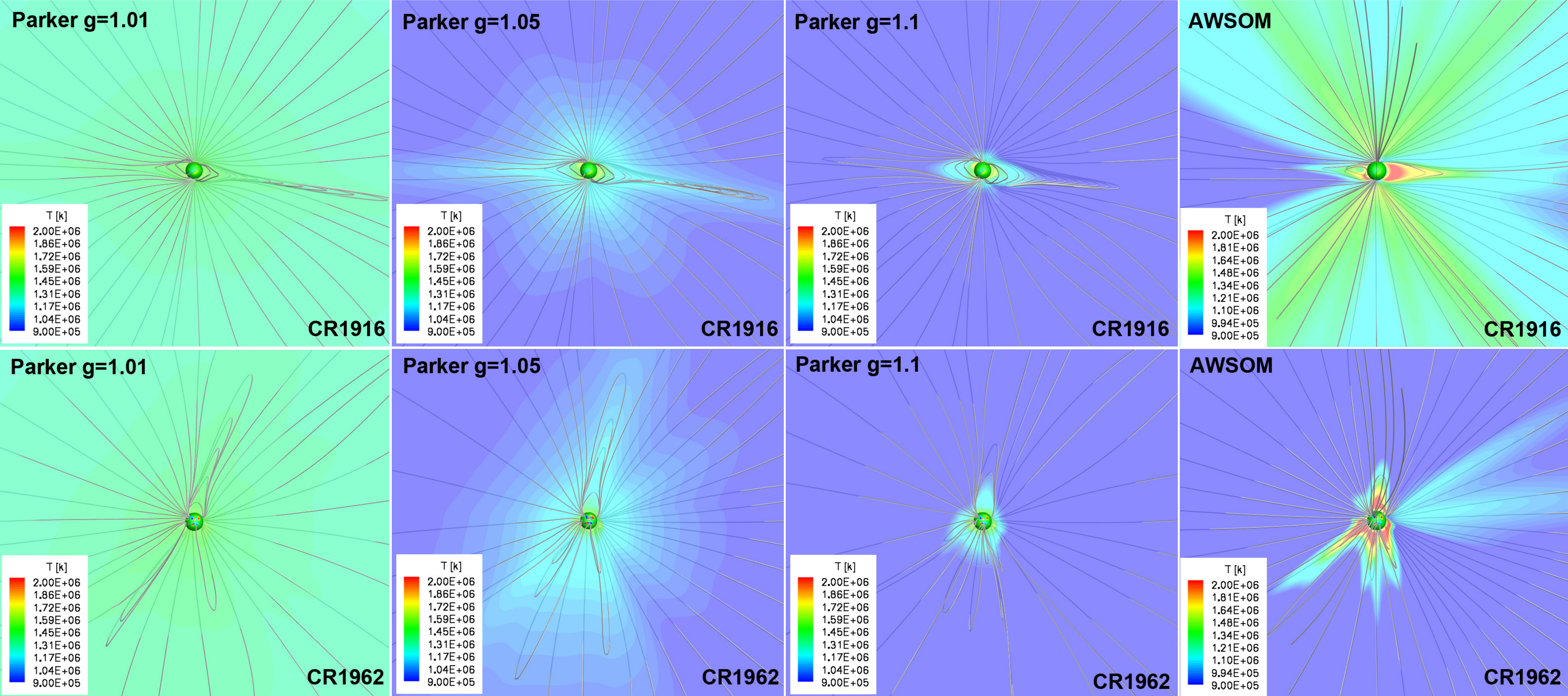}
\caption{Solar wind radial speed (top two panels) and coronal temperature (bottom two panels) solutions for CR1922 and CR1962 using the Parker model with three values of $\gamma$ (first three columns), and using the AWSOM model (right column). Also shown are selected field lines and the Alfv\'en surfaces (at the top two panels).}
\label{fig:f2}
\end{figure*}

Figure~\ref{fig:f2} shows the results for the wind speed and temperature from the polytropic model with $\gamma=1.01,1.05,1.1$, and from AWSOM. The wind speed in the polytropic solutions is quite uniform, with speeds ranging between $100\;km\;s^{-1}$ to about $500\;km\;s^{-1}$ for $\gamma=1.01$, $100\;km\;s^{-1}$ to about $400\;km\;s^{-1}$ for $\gamma=1.05$, and $100\;km\;s^{-1}$ to about $300\;km\;s^{-1}$ for $\gamma=1.1$. The fast/slow wind contrast is much more visible in the AWSOM solutions, with clear regions of slow wind with speed of about $300-400\;km\;s^{-1}$, and regions of fast wind with speed above $600\;km\;s^{-1}$.

The MHD solutions presented here use single fluid plasma, which assumes that the ion temperature, $T_i$, and the electron temperature, $T_e$, are equal, and that the plasma temperature, $T_p=(T_i+T_e)/2=p/2\rho$ (where $p$ and $\rho$ are the simulated pressure and density, respectively). The plasma temperature is half the prescribed temperature in the initial Parker wind and the coronal base temperature. Here we show this reduced temperature as it represents a more realistic coronal temperature (especially in the context of EUV/X-ray observations described below).

The temperature plots show, as expected from an isothermal solution, an almost uniform coronal temperature of 1.5MK for the $\gamma=1.01$ cases. For the $\gamma=1.05$ and $\gamma=1.1$ cases, the temperature contrast is more notable and it follows the density and magnetic field structure. In particular, the temperature contrast clearly follows the helmet streamers (the close field regions) in the solar minimum case. In the AWSOM solutions, the temperature contrast is very clear, and the temperature is much higher (over 2MK) in the helmet streamers. The temperature does not exceed the prescribed, 1.5MK in the polytropic solutions. 

\begin{figure*}[h!]
\centering
\includegraphics[width=5.5in]{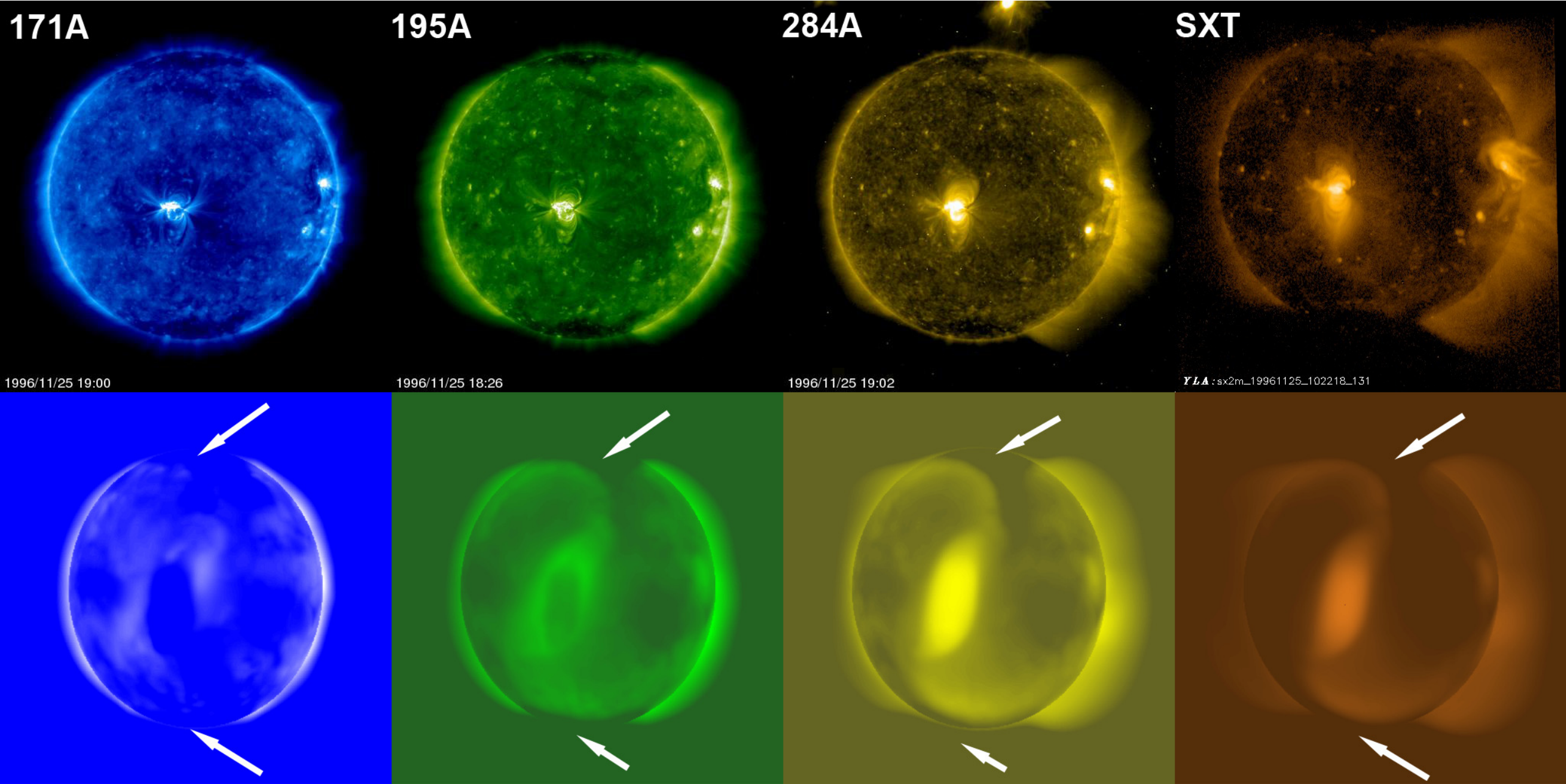}\\
\includegraphics[width=5.5in]{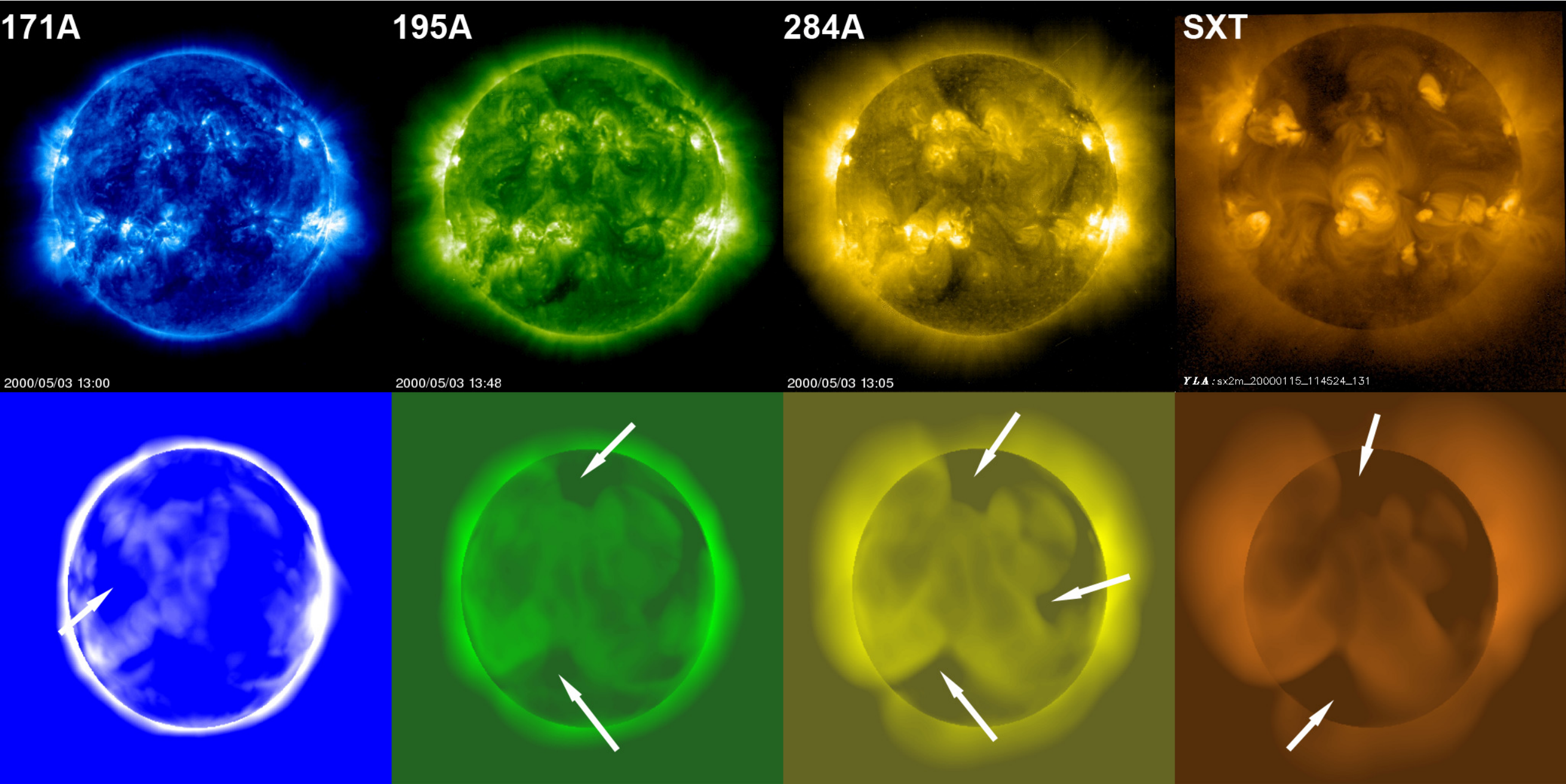}
\caption{SOHO EIT (171\AA, 195\AA, and 284\AA) and YOHKOH SXT images compared with synthetic images produced from the AWSOM solutions for CR1916 (top two panels) and for CR1962 (bottom two panels). The white arrows mark the location of the coronal holes in the synthetic images. Similar images produced from the Parker wind solutions are blank and do not show any features.}
\label{fig:f3}
\end{figure*}

Figure~\ref{fig:f3} shows a comparison between SOHO Extreme ultraviolet Imagine Telescope (EIT)\footnote
{{\tt http://sohodata.nascom.nasa.gov}} and YOHKOH Soft X-ray Telescope (SXT)\footnote
{{\tt http://ylstone.physics.montana.edu/}} full-disk images of the Sun, and synthetic images produced by AWSOM. The EIT bands are $171$, $195$, and $284$ {\AA}, and the X-ray images are integrated over the $2.4-32$ {\AA} range. The synthetic images are obtained by integrating the square of the electron density, $n_e$, multiplied by a response function for a particular temperature bin, $\Lambda(T)$ along the line-of-sight (LOS) through the three-dimensional solution. The response functions are calculated from the {\it CHIANTI} atomic database \citep[see e.g.,][]{Dare97,Landi13} in a similar way that the actual observed images are produced, taking into account the particular instrument calibration \citep[see][for more details about the production of synthetic LOS images in {\it BATS-R-US}]{Downs10,Oran15}. 

The comparison in Figure~\ref{fig:f3} shows a very good agreement between the observations and the images produced from AWSOM. The location and overall structure of the active regions is well reproduced in the modeled images, as well as the location and size of the coronal holes (the dark regions associated with the lower density in the open field regions). In the solar minimum case, the active region at the center of the disk is reproduced, along with the active region close to the right limb. The active region in the center is blank in the 171 {\AA} band, probably because the model overheated the area with temperature above the one, which responses to that particular line. For solar maximum, the overall structure of the active regions belts is reasonably reproduced, along with the coronal hole boundaries. We performed similar procedure to produce synthetic EIT/SXT images from the polytropic models. However, due to the lower coronal plasma temperature and the low temperature contrast all the LOS images are blank and do not show any features.  

The usage of ZDI maps to drive MHD models for the coronae and winds of cool stars is important in the context of stellar spindown and stellar evolution. Table~\ref{table:t1} summarizes the maximum wind speed, the total mass-loss rate, and total angular momentum loss rate of each of the solutions. 

\section{DISCUSSION}
\label{sec:Discussion}

The values from Table~\ref{table:t1} show that the polytropic models cannot produce consistent agreement with the observed properties of the Sun. The solutions with $\gamma=1.05,1.1$ cannot produce the fast (above $600\;km\;s^{-1}$) wind, while the $\gamma=1.01$ solution does produce a faster wind, but it consistently overestimates the observed solar mass-loss rate of about $\dot{M}_\odot\approx 2-3\cdot 10^{-14}\;M_\odot\;yr^{-1}$ \citep{Cohen11}. 
Reducing the boundary density of the $\gamma=1.05$ solution from $10^8$ to $5\cdot 10^7\;[cm^{-3}]$ does not produce fast wind as well, while the overestimation of the mass loss rate is reduced. Reducing the boundary density even further to $10^7\;[cm^{-3}]$ leads to a much faster wind, but too low mass-loss rate of $0.4\;10^{-14}\;M_\odot\;yr^{-1}$. The AWSOM solution produces fast winds above $600\;km\;s^{-1}$, and mass-loss rates of $1.3-1.6\;10^{-14}\;M_\odot\;yr^{-1}$, within a factor of 2 or so from the observed one. The two AWSOM solutions for the two solar epochs produce consistent values for the mass-loss and angular momentum loss rates, while the polytropic solutions slightly differ between solar minimum and solar maximum. Surprisingly, the polytropic solutions overestimate the mass-loss rate but underestimate the angular momentum loss rate for $\gamma=1.01,1.05$ comparing to the AWSOM solutions. The mass-loss rate is comparable between the $\gamma=1.1$ and the AWSOM solutions, but the angular momentum loss rate is overestimated by the $\gamma=1.1$ solution. 

The trends mentioned above can be explained by the different in size of the Alfv\'en surfaces in the different solutions (shown in Figure~\ref{fig:f2}). For the $\gamma=1.01,1.05$, the Alfv\'en surface is smaller than that in the AWSOM solution. Thus, the overall mass-loss is higher due to the higher density on the surface (closer to the solar surface), but the angular momentum loss rate is smaller due to the shorter lever arm that applies a torque on the star. The Alfv\'en surface is slightly bigger in the $\gamma=1.1$ solutions comparing to the AWSOM solutions. This explains the similar mass-loss rate but slightly larger angular momentum loss rate. Nevertheless, the wind speeds and temperatures of the $\gamma=1.1$ solutions do not agree with solar observations.

\begin{figure*}[h!]
\centering
\includegraphics[width=5.5in]{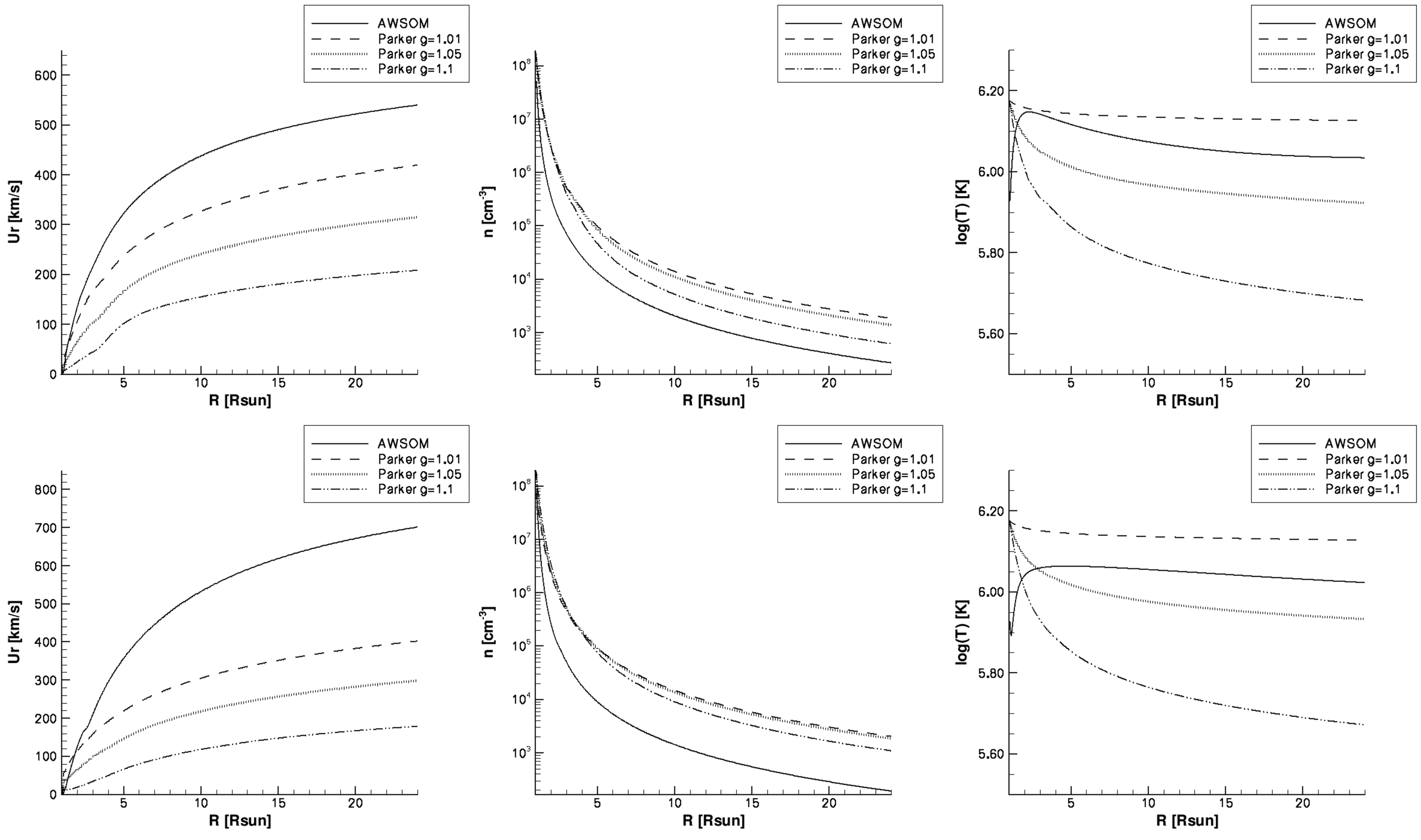}
\caption{The wind's radial speed (left), number density (middle), and temperature (right) extracted along an open filed line with the same footpoint from the different solutions for solar minimum (top) and solar maximum (bottom).}
\label{fig:f4}
\end{figure*}

\begin{figure*}[h!]
\centering
\includegraphics[width=4.5in]{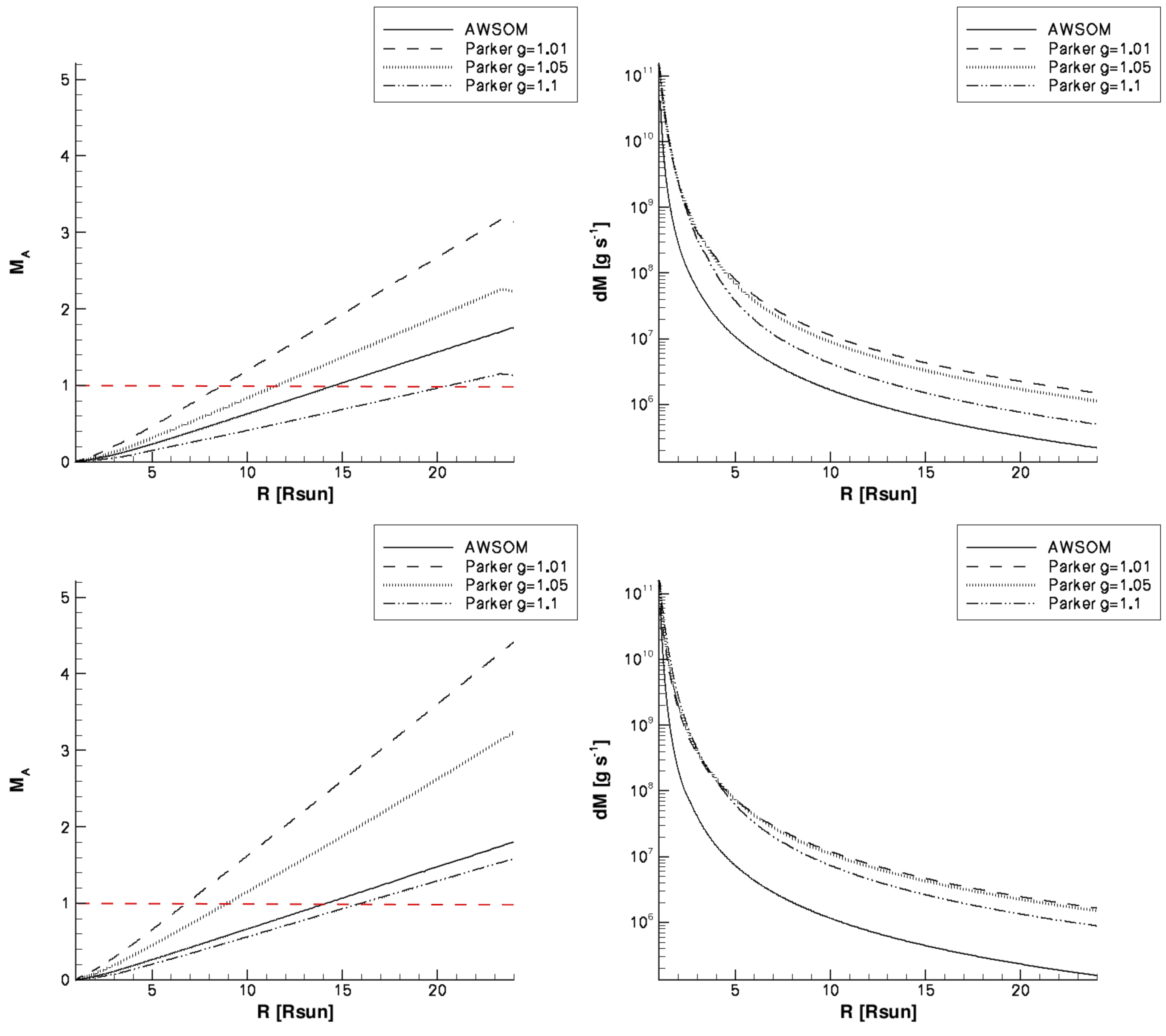}
\caption{The wind's Alfv\'enic Mach number, $M_A$ (left), and local mass loss rate, $dM$ (right) extracted along an open filed line with the same footpoint from the different solutions for solar minimum (top) and solar maximum (bottom). Dashed red line marks the $M_A=1$ line.}
\label{fig:f5}
\end{figure*}

Explanation for these trends can also be found in Figure~\ref{fig:f4}, where we show the wind's radial speed, number density, and temperature extracted along an open field lines with the same footpoint location from all the different solutions. Figure~\ref{fig:f4} shows that the wind is accelerated much faster in the AWSOM solution to higher values, where the polytropic speeds do not exceed $400\;km\;s^{-1}$. Only in the AWSOM solution, the temperature first rises and then falls adiabatically, while it only falls in the polytropic solutions. Finally, the density drops much faster in the AWSOM solutions comparing to the polytropic solutions. Figure~\ref{fig:f5} shows a similar extraction along the same field lines for the Alfv\'en Mach number, $M_A$, and the local mass-loss rate, $dM_i=\rho_i u_i 4\pi r^2_i$, where the index $i$ stands for the particular point along the field line. The plots show that the wind exceeds the Alfv\'en speed in the polytropic solutions with $\gamma=1.01,1.05$ $4-5$ solar radii lower than in the AWSOM solutions, where $M_A=1$ around $r=14-15R_\odot$. The wind exceeds the Alfv\'en speed at higher radii for the $\gamma=1.1$ polytropic solution comparing to the AWSOM solution. The local mass-loss rate trends are similar to those of the density, where $dM$ is 5-10 times higher at the location of $M_A=1$ for the polytropic solutions comparing to the AWSOM solution. This explains the overestimation of the mass-loss rate for $\gamma=1.01,1.05$ comparing to AWSOM. This trend of slower decline of the local mass-loss rate is compensated by the further Alfv\'en point in the $\gamma=1.1$ solution. Therefore, this solution does not overestimate the mass-loss rate of the AWSOM solution as it accounts for lower mass-loss rate at further distance. 

\cite{LeerHolzer80} found that if the energy deposition occurs below the Alfv\'en point it affects mostly the mass flux, without affecting much the final wind speed. Alternatively, they found that energy deposition above the Alfv\'en point affects mostly the final wind speed, without affecting much the mass flux. In our simulations, the Alfv\'en point seems to be closer to the surface in the polytroic model comparing to the AWSOM model. This result might be due to the fact that the polytropic model is set uniformly, and it is driven and constrained exclusively by the boundary conditions. Thus, the polytropic model is set by a small number of {\it global} constrains. The AWSOM model on the other hand, accounts for the heating and acceleration at each point of the domain individually, and the sources and sinks of energy and momentum are defined in a {\it local} manner, while also taking into account more detailed processes. This difference seems to make a significant difference on the location of the Alfv\'en point. Following \cite{LeerHolzer80}, it is possible that since the Alfv\'en point in the polytropic is rather low, the mass flux cannot be regulated much above it, resulting in an overestimation of the mass loss rate. The resulting relatively high density above the Alfv\'en point (comparing to the AWSOM density at similar heights) prevents the acceleration of the plasma to fast speeds as obtained by the AWSOM model.

The results show that the AWSOM model agrees with all elements of the observed solar properties -- the wind speed, coronal temperature and density structure, and total mass-loss rate (the total observed angular momentum loss rate is harder to determine). Non of the polytropic solutions can provide such a consistency with solar observations, and some of the solutions might even be considered "unphysical", due to a maximum wind speed that is lower than the minimum observed solar wind speed. The choice of the coronal base number density, $n$, in the models is crucial to get both agreement with the observed total mass-loss rate, and with the observed three-dimensional coronal density. Our work suggest that this base density should be about $n=10^8\;cm^{-3}$, which is lower than the choice in other polytropic models for stellar coronae \citep[e.g.,][]{Vidotto15,Reville15}. 

\begin{figure*}[h!]
\centering
\includegraphics[width=5.5in]{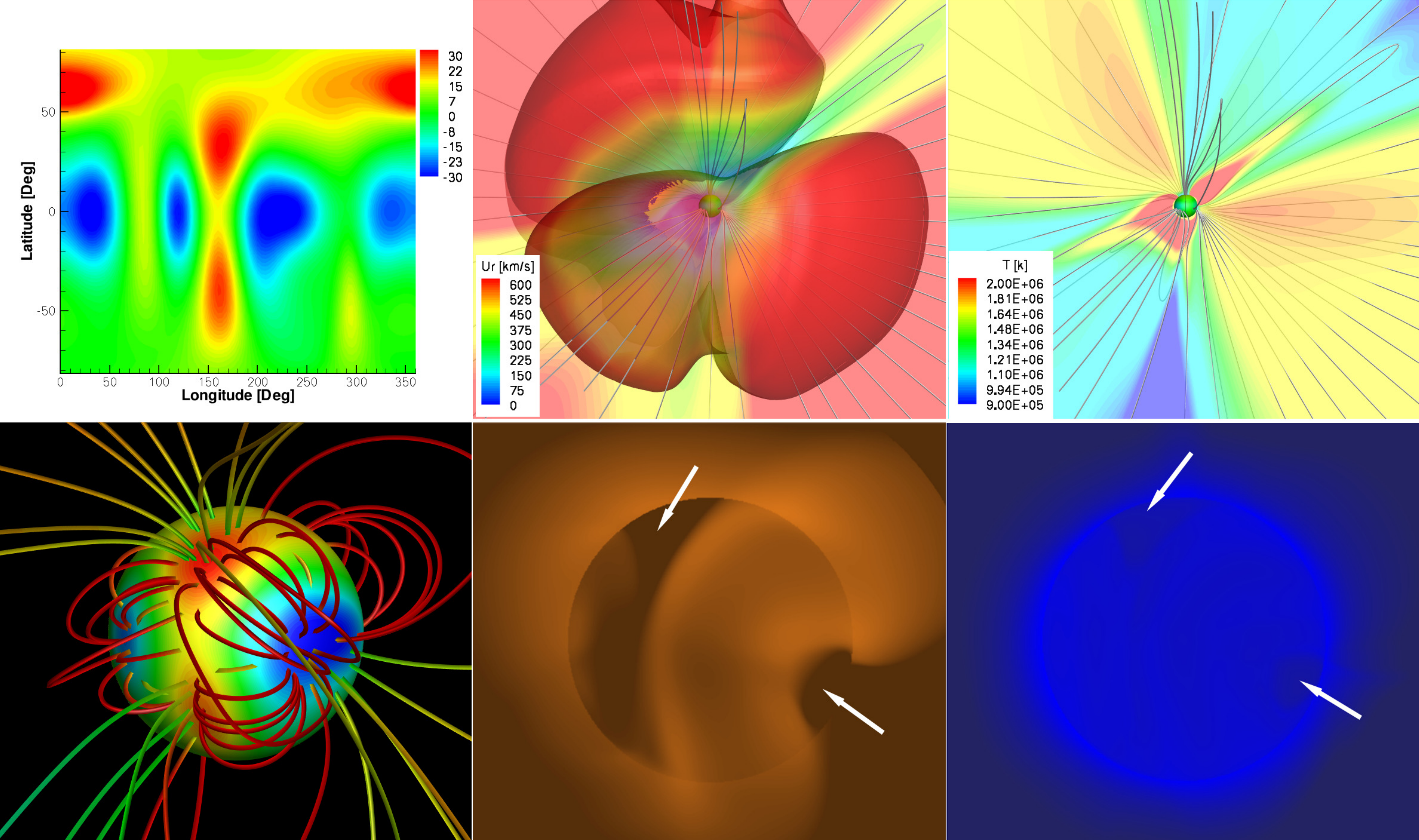}
\caption{Top: the surface radial magnetic field of HD18733 used here to drive the model (left), along with the AWSOM HD189733 solutions for the wind speed (middle) and the temperature (right) displayed in the same manner as in Figure~\ref{fig:f2}. Bottom: the three-dimensional coronal field of HD189733 (left) colored with temperature contour of hotter (red) and colder (green/yellow) plasma, along with EIT (middle), and SXT (right) synthetic images of HD189733. White arrows mark the coronal holes associated with the open, colder field lines.}
\label{fig:f6}
\end{figure*}

In order to extend this work to the stellar context, we perform similar simulations to the planet-hosting star HD 189733 using ZDI map which is taken from the published data by \cite{Fares10}. In the simulation of HD189733, we use the same parameter setting as in the solar runs. Figure~\ref{fig:f6} shows the solution for HD189733, which shows overall similar range of wind speeds and coronal temperatures comparing to the solar results. The figure also shows the three-dimensional coronal magnetic field, colored with temperature contours with hotter closed field lines (shown in red) and colder open filed lines (shown in green/yellow). We also show synthetic EIT/SXT images of HD189733, where a good agreement is obtained between the location of the colder open field lines and the coronal holes in the LOS images. Table~\ref{table:t2} shows the global parameters of HD189733 using the $\gamma=1.01$ polytropic model and AWSOM. The mass-loss rate is overestimated by the polytropic model comparing to AWSOM, but the angular momentum loss rate is comparable. However, just like in the solar case, the polytropic model cannot produce a fast wind. The mass-loss rate for HD189733 is similar to the solar one in the AWSOM model, but the angular momentum loss rate is much higher, probably due to the difference in rotation period (about 12 days comparing to the solar 25 days rotation period).

\section{Conclusions}
\label{sec:Conclusions}

We perform a comparative study between a polytropic wind MHD model and the physics-based AWSOM in order to test which model agrees better with solar observations. The AWSOM model produces a clear bi-model solar wind, with fast wind above $600\;km\;s^{-1}$, a good agreement (within a factor of 2) with the observed total solar mass-loss rate, and a good agreement with the coronal temperature and density structure (validated by solar full-disk LOS observations). The polytropic wind model, with three choices of the polytropic index, $\gamma$, fails to produce solutions that are consistent with solar observations. In particular, the polytropic model cannot produce the fast solar wind speed, and the coronal density drops too slow in this type of model. This leads to inconsistency with the observed total mass-loss rate, solar wind speed, and the coronal base density. 

Our recommendation is that polytropic models, which are used to estimate mass-loss rates and other parameters of solar analogs, must be first calibrated with solar observations. Alternatively, these models can be calibrated with models that capture more detailed physics of the solar corona (such as the AWSOM model), and that can reproduce solar observations in a consistent manner. Without such a calibration, the results of the polytropic models cannot be validated, but they can be wrongly used by others. 


\acknowledgments

We thank an unknown referee for his/her comments. The work presented here was funded by a NASA Living with a Star grant number NNX16AC11G. Simulation results were obtained using the Space Weather Modeling Framework, developed by the Center for Space Environment Modeling, at the University of Michigan with funding support from NASA ESS, NASA ESTO-CT, NSF KDI, and DoD MURI. The simulations were performed on the Smithsonian Institute HYDRA cluster. 




\begin{table*}[h!]
\caption {Simulations Global Parameters of the Solar Corona} \label{tab:table1}
\resizebox{1.\textwidth}{!}{\centering\begin{tabular}{ccccccccccc}
\hline
{\bf Case}&{\bf CR1916 Parker} & {\bf CR1916 Parker} &  {\bf CR1916 Parker} &  {\bf CR1962 Parker} & {\bf CR1962 Parker} &{\bf CR1962 Parker}&{\bf CR1916 Parker} &{\bf CR1916 Parker} &{\bf CR1916 AWSOM}&{\bf CR1962 AWSOM}\\
&{\bf $\gamma=1.01$}&{\bf $\gamma=1.05$}&{\bf $\gamma=1.1$}&{\bf $\gamma=1.01$}&{\bf $\gamma=1.05$}&{\bf $\gamma=1.1$}&{\bf $\gamma=1.05$ $n=10^7$}&{\bf $\gamma=1.05$ $n=5\cdot 10^7$}& & \\
\hline
Maximum Speed $[km/s]$&460&350&275&518&440&520&1400&367&680&770\\
$\dot{M}\;[10^{-14}\;M_\odot/yr]$&9.1&5.3&1.5&6.8&5.2&0.9&0.4&2.0&1.3&1.6\\
$\dot{J}\;[10^{28}\;g\;cm^2/s^2]$&9.5&5.5&1.8&6.8&5.2&2.4&0.5&2.1&1.1&2.7\\
\hline
\end{tabular}}
\label{table:t1}
\end{table*}

\begin{table*}[h!]
\caption {Simulations Global Parameters of HD 189733} \label{tab:table2}
\centering
\begin{tabular}{ccc}
\hline
{\bf Case}&{\bf Parker $\gamma=1.01$} & {\bf AWSOM} \\
\hline
Maximum Speed $[km/s]$&500&946\\
$\dot{M}\;[10^{-14}\;M_\odot/yr]$&4.1&1.6\\
$\dot{J}\;[10^{28}\;g\;cm^2/s^2]$&4.1&1.7\\
\hline
\end{tabular}
\label{table:t2}
\end{table*}


\end{document}